# Decentralized set-valued state estimation and prediction for hybrid systems: A symbolic approach

N. Bajcinca and Y. Kouhi

*Abstract*— A symbolic approach to decentralized set-valued state estimation and prediction for systems that admit a hybrid state machine representations is proposed. The decentralized computational scheme represents a conjunction of a finite number of distributed state machines, which are specified by an appropriate decomposition of the external signal space. It aims at a distribution of computational tasks into smaller ones, allocated to individual distributed state machines, leading to a potentially significant reduction in the overall space/time computational complexity. We show that, in general, such a scheme outerapproximates the state set estimates and predictions of the original monolithic state machine. By utilizing structural properties of the transition relation of the latter, in a next step, we propose constructive decomposition algorithms for a recovery of the exact state set outcomes.

## I. Introduction

Conservative set-valued based computation has versatile applications in the analysis and synthesis of complex systems. In particular, such an approach can be efficiently employed for the prediction of the system's behaviour comprehending physical and measurement uncertainties. Its usability is justified as the state set estimates are guaranteed to contain the true system state. However, the estimation computational cost increases, in general, exponentially with the system's state space dimension. Therefore, the decentralized approach, leading to a potential reduction in computational complexity, has received a considerable attention, particularly in discrete-event systems (see e.g. [1] and its references).

This article follows a similar approach in dealing with the state estimation problem using hybrid state machine representations with a finite external signal space. Such symbolic models are obtained by a discrete approximation of the behaviour of continuous, discrete-event and hybrid systems, see e.g. [2] and the references therein. For the purpose of decentralized computation, the signal space is decomposed into a finite number of subspaces, equipped with specific aggregation functions. It turns out that, in general, the decentralized scheme provides conservative overapproximate outcomes as compared to the monolithic ones. For a recovery of the exact state set estimates and predictions, constructive decomposition algorithms by utilizing the structural properties of the transition relation of the monolithic state machine are devised. Therefore the simple concept of "non-deterministic chains" – first introduced in [3] – has been employed. Non-deterministic chains represent a special class of transition relations featuring inherent injectivity properties in the corresponding transition functions. We show that for every transition relation that assumes a partitioning into a set of non-deterministic chains, conjunctive decentralized schemes producing exact state set estimation exist always. Moreover, by means of proper state space aggregations, we conceive an algorithm which extends this simple idea to the general class with non-injective state set transition functions.

The remaining of the paper is organized as follows. In Section II the reader is made familiar with the used notation and basic preliminary concepts. Section III represents the core of the work. We define here the conjunctive decentralized computation scheme, and introduce algorithms for the decomposition of the external signal space. Several examples are used to illustrate the main ideas and procedures.

## II. Preliminaries

### A. Notation & conventions

Capital letters denote signal spaces, e.g. $X$ and $W$ represent the state space and the external signal space, respectively. The corresponding elements of a space are denoted by greek lowercase letters, e.g. $W = \{\omega_1, \ldots, \omega_m\}$, $\xi \in X$, etc. We consider the discrete time domain, hence signals, which are denoted by lowercase letters, are sequences of symbols from the appropriate signal space, e.g. $w: \mathbb{N}_0 \to W$ represents an external signal. The restriction of a signal, or a set of signals to an interval $[\tau, t]$, with $0 \leq \tau \leq t \in \mathbb{N}_0$, is denoted by $\cdot|_{[\tau,t]}$, e.g. $w|_{[\tau,t]} = w(\tau)w(\tau+1)\ldots w(t)$. The space of the finite sequences (strings) $w|_{[\tau,t]}$ will be denoted as $W^{[\tau,t]} = W^{t-\tau+1}$. The string $w|_{[\tau,t]}$ will be considered as an element of $W^{[\tau,t]}$, i.e. they will be represented by an $(t-\tau+1)$-tuple ordered by the time parameter.

Let $\sim$ be an equivalence relation defined on a set $W$. The equivalence class of an element $\alpha \in W$ is the subset of all elements in $W$ which are equivalent to $\alpha$, that is: $[\alpha] = \{\omega \in W; \omega \sim \alpha\}$. The quotient set of $W$ by $\sim$ is defined as $W/\sim = \{[\alpha]; \alpha \in W\}$. The canonical projection map $\pi: W \to W/\sim$ maps elements of $W$ to equivalence classes. Finally, the equivalence classes will be labeled by symbols with the help of an injective function $L: W/\sim \to V = \{\theta_1, \ldots, \theta_q\}$, which we refer to as a labeling function.

Let $f$ be an arbitrary function $f$ defined on some domain $X$. If $\Xi$ is a subset of $X$, we will use the convention $f(\Xi) := \cup_{x \in \Xi} f(x)$. For singletons we often avoid the brackets, that is instead of $\{\omega\}$ we rather use $\omega$.

## B. Systems & realizations

A dynamical system $\Sigma$ is defined as a triple $(T, W, \mathcal{B})$, with *time axis* $T \subseteq \mathbb{R}$, the *external signal space* $W$, and the *behaviour* $\mathcal{B} \subseteq W^T$, where $W^T = \{w : T \to W\}$, see [4]. In words, $\mathcal{B}$ represents a family of sequences $w : \mathbb{N}_0 \to W$ which are compatible with the dynamics of the system $\Sigma$. A *state machine* is defined as a tuple $P = (X, W, \Delta, X_0)$ where $X$ denotes the state space, $W$ the external signal space, $\Delta \subseteq X \times W \times X$ the transition relation, and $X_0 \subseteq X$ the initial state set. If $X = \mathbb{R}^n \times D$, where $n \in \mathbb{N}$ and $D \subset \mathbb{N}$ is a finite set, then $P$ is referred to as a *hybrid state machine*; for $n = 0$, $P$ is a *finite state machine*. For the sake of simplicity, here $P$ is assumed generally to be non-blocking and $X_0 = X$.

Hybrid state machines cover a wide range of system classes including the time-driven (continuous), event-driven and hybrid systems. The synthesis of the transition relation $\Delta$ consists in symbolic encoding of the system behaviour $\mathcal{B}$ in terms of the transitions between the states in $X$. For illustration purposes, consider a time-driven continuous system defined by $\dot{\xi} = a(\xi)$ where $\xi \in X \subseteq \mathbb{R}^n$ and $a : X \to X$. Introduce the external signal space $W = L(X/Q)$, where $Q$ represents a finite equivalence relation in $X$, and $L$ a labeling function. Unique solutions $\phi(t, \xi)$ can be associated with each initial value $\xi \in X$ if $a : X \to X$ is Lipschitz on $X$. Then, $(\xi, \omega, \xi') \in \Delta$ if $\xi' = \phi(T_s, \xi)$ with $T_s$ representing the sampling time, and $L(\pi_Q(\xi')) \neq L(\pi_Q(\xi)) = \omega \in W$. If $\phi(t, \xi) \in \pi_Q(\xi)$ for all $t \in \mathbb{R}$, then we adopt $(\xi, \omega, \xi) \in \Delta$. Other encoding scenarios for the transition relation $\Delta$ can be utilized alternatively.

For systems exhibiting an input/output structure, the external signal space $W$ can be decomposed as $W = U \times Y$, with $U$ and $Y$ being the sets of input and output symbols. Then $P = (X, U \times Y, \Delta, X_0)$ is said to be an *I/S/- machine* if for each reachable state $\xi \in X$ and each $\mu \in U$, there exists a $\nu \in Y$ and a $\xi' \in X$, such that $(\xi, (\mu, \nu), \xi') \in \Delta$. If $\nu$ and $\xi'$ are unique for all $\xi \in X$ and $\mu \in U$, $P$ is said to be an *I/S/O machine*. Note that I/S/O machines are by definition deterministic, that is, if $(\xi, (\mu, \nu), \xi') \in \Delta$ and $(\xi, (\mu, \nu), \xi'') \in \Delta$, then $\xi' = \xi''$.

A state machine $P = (X, W, \Delta, X_0)$ induces a state space system $\Sigma_S = (\mathbb{N}_0, W \times X, \mathcal{B}_S)$, where $\mathcal{B}_S$ is referred to as the *full behaviour*, and is defined as

$$\mathcal{B}_S := \{(w, x); (x(t), w(t), x(t+1)) \in \Delta, t \in \mathbb{N}_0, x_0 \in X_0\}. \quad (1)$$

The *external behaviour* $\mathcal{B}_{\text{ex}}$ of $\Sigma_S$ is then defined to be the projection of $\mathcal{B}_S$ onto $W^{\mathbb{N}_0}$, that is $\mathcal{B}_{\text{ex}} := \mathcal{P}_W \mathcal{B}_S = \{w; \exists x \in X^{\mathbb{N}_0}, (w, x) \in \mathcal{B}_S\}$. A *state space path* associated with a string $w|_{[\tau,t]}$, denoted as $(\xi_\tau, w|_{[\tau,t]}, \xi_{t+1})$, is said to be *present* in the state machine $P$, if $(w, x) \in \mathcal{B}_S$ exists such that $\xi_\tau = x(\tau)$ and $\xi_{t+1} = x(t+1)$. Finally, a state machine $P = (X, W, \Delta, X_0)$ with induced external behaviour $\mathcal{B}_{\text{ex}}$ is called a realization of a dynamical system $\Sigma = (\mathbb{N}_0, W, \mathcal{B})$ if $\mathcal{B}_{\text{ex}} = \mathcal{B}$. This will be denoted by $P \cong \Sigma$.

## C. State sets

Let $\mathcal{B}_S$ and $\mathcal{B}_{\text{ex}}$ be the induced full and external behaviour of the state machine $P = (X, W, \Delta, X)$, respectively, and consider a string $w|_{[\tau,t]} \in W^{[\tau,t]}$. The sets defined by

$$\chi(w|_{[\tau,t]}) := \{\xi; \exists (w', x) \in \mathcal{B}_S, x(t) = \xi, w'|_{[\tau,t]} = w|_{[\tau,t]}\}, \quad (2a)$$
$$\rho(w|_{[\tau,t]}) := \{\xi; \exists (w', x) \in \mathcal{B}_S, x(t+1) = \xi, w'|_{[\tau,t]} = w|_{[\tau,t]}\}, \quad (2b)$$

will be referred to as the *estimated* and *predicted* state sets compatible with the external finite sequence $w|_{[\tau,t]}$, respectively. Both, $\chi$ and $\rho$, represent families of set-valued functions $W^{[\tau,t]} \to 2^X$. Note that $w|_{[\tau,t]} \in \mathcal{B}_{\text{ex}}|_{[\tau,t]} \Leftrightarrow \chi(w|_{[\tau,t]}) \neq \emptyset$ and $\rho(w|_{[\tau,t]}) \neq \emptyset$.

In general, more information from the past leads to more accurate state estimations and predictions. This fact is reflected by the following general inclusion relationships

$$\chi(w|_{[0,t]}) \subseteq \chi(w|_{[1,t]}) \subseteq \ldots \subseteq \chi(w|_{[t,t]}), \quad (3a)$$
$$\rho(w|_{[0,t]}) \subseteq \rho(w|_{[1,t]}) \subseteq \ldots \subseteq \rho(w|_{[t,t]}). \quad (3b)$$

Introduce, further, the *parametrized state transition* function $\hat{\rho}_\omega : X \to 2^X$, where $\omega \in W$, with

$$\hat{\rho}_\omega(\xi) := \{\xi'; (\xi, \omega, \xi') \in \Delta\}. \quad (4a)$$

For $\Omega \subseteq W$, in accordance with the adopted convention

$$\hat{\rho}_\Omega(\Xi) := \cup_{\omega \in \Omega, \xi \in \Xi} \hat{\rho}_\omega(\xi). \quad (4b)$$

Then, the predicted states in $X$, resulting from the occurrence of the symbol $\omega \in W$, are computed by

$$\rho(\omega) = \hat{\rho}_\omega(\chi(\omega)), \quad (4c)$$

where according to (2a)

$$\chi(\omega) = \{\xi; \exists \xi', (\xi, \omega, \xi') \in \Delta\}, \quad (4d)$$

with $\omega = w|_{[t,t]}$. For sequences,

$$\rho(w|_{[\tau,t]}) = \hat{\rho}_{w(t)}(\chi(w|_{[\tau,t]})). \quad (4e)$$

Notice, that in light of this equation, (3b) follows directly from (3a) as the set in the argument of $\hat{\rho}_{w(t)}$ is shrinking. Moreover, by definition (2a)

$$\chi(w|_{[\tau,t]}) = \rho(w|_{[\tau,t-1]}) \cap \chi(w(t)), \quad (4f)$$

which along with (4e) reveals a recursive structure in computing $\chi(w|_{[\tau,t]})$ and $\rho(w|_{[\tau,t]})$. Observe that now in view of (3b), equation (3a) follows immediately from (4f).

## III. DECENTRALIZED COMPUTATION

### A. Signal space decomposition

Introduce a finite set of equivalence relations $A_k$ on the external signal space $W$, with $\pi_k$ representing the corresponding canonical projection map

$$\pi_k : W \to W/A_k, \quad k \in \{1, \ldots, p\}. \quad (5)$$

Introduce functions $L_k : W/A_k \to V_k$ which assign to each equivalence class $[\omega]_{A_k}$ a labeling symbol $\theta_k$. The composition $\mathcal{A}_k := L_k \circ A_k$,

$$\mathcal{A}_k : W \to V_k, \quad k \in \{1, \ldots, p\}, \quad (6)$$

ushers $V_k$ as an "aggregate" space of the external signal space $W$. Formally, we denote this by $V_k = L_k(W/A_k)$.

Throughout the article we consider classes of equivalence relations $A_k$, $k \in \{1,\ldots,p\}$, which fulfill the following *consistency* condition: $\cap_{k=1}^p [\omega]_{A_k} = \omega, \forall \omega \in W$. With regard to (6), this is equivalently stated by

$$\cap_{k=1}^p \mathcal{A}_k^{-1}(\mathcal{A}_k(\omega)) = \omega \quad (\forall \omega \in W). \tag{7}$$

Then, each symbol $\omega \in W$ is uniquely associated with an ordered $p$-tuple of symbols $(\theta_1,\ldots,\theta_p) \in V_1 \times \ldots \times V_p$. This will be referred to as the *decomposition* of the signal space $W$ and formally designated as

$$W \rightsquigarrow V_1 \times \ldots \times V_p. \tag{8}$$

Note that in general the opposite does not hold: not every $p$-tuple in $V_1 \times \ldots \times V_p$ has an associate in $W$. Throughout the article we will adhere to the consistency condition (7).

Now consider the space of infinite sequences $W^{\mathbb{N}_0}$, and introduce the equivalence relation $A_k$ thereon as: $(w',w'') \in A_k$ if $(w'(t),w''(t)) \in A_k$, for all $t \in \mathbb{N}_0$. In this sense, each sequence $w \in \mathbb{N}_0$ is assigned an equivalence class $[w]_{A_k} := \prod_t [w(t)]_{A_k} = [w(0)]_{A_k} \times [w(1)]_{A_k} \times \ldots$ whereby in accordance with the adopted conventions in Section II-A, the sequences in $W^{\mathbb{N}_0}$ are represented by $\infty$-tuples. A natural labeling policy for such equivalence classes in $W^{\mathbb{N}_0}/A_k$ by making use of those for $W/A_k$ leads to labels in form of sequences in $V_k^{\mathbb{N}_0}$: $L_k([w]_{A_k}) := L_k([w(0)]_{A_k}) L_k([w(1)]_{A_k}) \ldots = \theta_k^0 \theta_k^1 \ldots$, where $\theta_k^t = L_k([w(t)]_{A_k})$, $\forall t \in \mathbb{N}_0$. This infers the labeling map $L_k := W^{\mathbb{N}_0}/A_k \to V_k^{\mathbb{N}_0}$, which leads to the extension of (6) to $\mathcal{A}_k : W^{\mathbb{N}_0} \to V_k^{\mathbb{N}_0}$, $k \in \{1,\ldots,p\}$ by $\mathcal{A}_k = L_k \circ A_k$. It is now obvious that if consistency condition (7) for symbols holds, so it holds for sequences, as well. Specifically,

$$\cap_{k=1}^p \mathcal{A}_k^{-1}(\mathcal{A}_k(w|_{[\tau,t]})) = w|_{[\tau,t]} \quad (w \in W^{\mathbb{N}_0}). \tag{9}$$

Then a string $w|_{[\tau,t]}$ is always associated with a unique tuple of strings $(v_1|_{[\tau,t]},\ldots,v_p|_{[\tau,t]}) \in V_1^{[\tau,t]} \times \ldots \times V_p^{[\tau,t]}$. It can be also shown that for any $w \in W^{\mathbb{N}_0}$ and $\mathcal{A}_k$

$$\mathcal{A}_k^{-1}(\mathcal{A}_k(w|_{[\tau,t]})) = \mathcal{A}_k^{-1}(\mathcal{A}_k(w))|_{[\tau,t]}. \tag{10}$$

Note also that the aggregation maps $\mathcal{A}_k$ associate with each $w|_{[\tau,t]} \in W^{[\tau,t]}$ a subset of $W^{[\tau,t]}$ defined by

$$\mathcal{D}_\mathcal{A}(w|_{[\tau,t]}) := \cup_{k=1}^p \mathcal{A}_k^{-1}(\mathcal{A}_k(w))|_{[\tau,t]}. \tag{11}$$

The behaviour of the map $\rho$ within such a domain for a fixed $w|_{[\tau,t]}$ will play a key role in our forthcoming derivations.

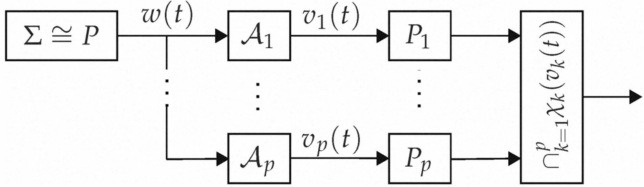

Fig. 1. Decentralized set-valued state estimation scheme.

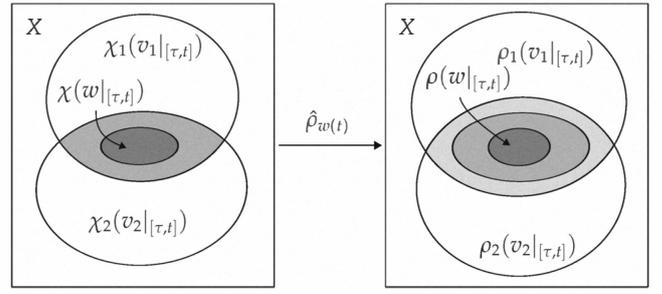

Fig. 2. Illustrating the decentralized set-valued state estimation and prediction in a setting with two state machines $P_1$ and $P_2$ exerting the strings $v_1|_{[\tau,t]}$ and $v_2|_{[\tau,t]}$, respectively, which are consistent to $w|_{[\tau,t]}$; refer to (14a) and (14b).

### B. Distributed state machines

The equivalence relations $A_k$, as defined in the previous section, bring us to the concept of *distributed state machines*.

*Definition 1:* Consider a monolithic state machine $P = (X,W,\Delta,X)$. The distributed state machines $P_k$, with $k \in \{1,\ldots,p\}$, induced by the equivalence relations $A_k$ on the external signal space $W$ are defined as $P_k = (X,V_k,\Delta_k,X)$, with $V_k = L_k(W/A_k)$, and $\Delta_k \subseteq X \times V_k \times X$ given by

$$\Delta_k = \{(\xi,\theta_k,\xi'); \exists \omega \in \mathcal{A}_k^{-1}(\theta_k), (\xi,\omega,\xi') \in \Delta\}. \tag{12}$$

Hereof it is clear that the full and external behaviour of a distributed state machine $P_k$ are determined by those of the monolithic state machine $P$: $\mathcal{B}_{s,k} = \{(\mathcal{A}_k(w),x); (w,x) \in \mathcal{B}_s\}$ and $\mathcal{B}_{ex,k} = \{\mathcal{A}_k(w); w \in \mathcal{B}_{ex}\} =: \mathcal{A}_k(\mathcal{B}_{ex})$, respectively. Clearly, $\cap_{k=1}^p \mathcal{A}_k^{-1}(\mathcal{A}_k(\mathcal{B}_{ex})) \supseteq \mathcal{B}_{ex}$.

The corresponding decentralized estimation and prediction functions $\chi_k : V_k^{[\tau,t]} \to 2^X$ and $\rho_k : V_k^{[\tau,t]} \to 2^X$ of the distributed state machines $P_k$ are defined using analogous expressions to (2a-2b). In fact, $\rho_k$ and $\chi_k$ can be computed as

$$\chi_k = \chi \circ \mathcal{A}_k^{-1}, \quad \rho_k = \rho \circ \mathcal{A}_k^{-1}. \tag{13}$$

If $\rho$ is applied on both sides of (9), then using the fact that for any function $\phi$: $\phi(M_1 \cap M_2) \subseteq \phi(M_1) \cap \phi(M_2)$ we get

$$\begin{aligned} \rho(w|_{[\tau,t]}) &= \rho\left(\cap_{k=1}^p \mathcal{A}_k^{-1}(\mathcal{A}_k(w|_{[\tau,t]}))\right) \\ &\subseteq \cap_{k=1}^p \rho\left(\mathcal{A}_k^{-1}(\mathcal{A}_k(w|_{[\tau,t]}))\right) \\ &= \cap_{k=1}^p \rho_k\left(v_k|_{[\tau,t]}\right), \end{aligned} \tag{14a}$$

where (13) is used in the 3$^{\text{rd}}$ line, and $v_k|_{[\tau,t]} = \mathcal{A}_k(w|_{[\tau,t]})$ represent the consistent external strings to $w|_{[\tau,t]}$, exerted on the distributed state machines $P_k$. Similarly,

$$\cap_{k=1}^p \chi_k(v_k|_{[\tau,t]}) \supseteq \chi(w|_{[\tau,t]}). \tag{14b}$$

Equations (14a) and (14b) suggest a decentralized computation scheme for the set-valued state estimate $\chi(w|_{[\tau,t]})$ and prediction $\rho(w|_{[\tau,t]})$ involving $p$ state machines $P_k$ as depicted in Fig. 1. Referring to (14a) and (14b), such a parallel computation provides in general overapproximate results of the monolithic machine $P$; see Fig. 2 for a mapping picture of the setup with $p = 2$. Yet for certain classes of transition relations $\Delta$ of the monolithic state machine $P$, specific decomposition policies (8) may lead to a decentralized scheme producing the outcomes that exactly match with those of the monolithic state machine $P$. To this end, our

effort will particularly consist in a suitable manipulation of the aggregation maps $\mathcal{A}_k$, or, equivalently, the signal space decomposition (8). Therefor, in this article, we employ the concept of "non-deterministic chains".

*C. Non-deterministic chains*

*Definition 2:* Consider a state machine $P = (X, W, \Delta, X)$. A transition subrelation $\delta \subseteq \Delta$ is said to be a *non-deterministic chain* over a subspace $\Omega \subseteq W$ if

$$\delta = \{(\xi, \omega, \xi') \in \Delta; \forall \omega \in \Omega\}, \tag{15}$$

and for all $\omega', \omega'' \in \Omega$:

(i) $(\xi, \omega', \xi') \in \delta, (\xi, \omega'', \xi'') \in \delta \Rightarrow \omega' = \omega''$,

(ii) $(\xi', \omega', \xi) \in \delta, (\xi'', \omega'', \xi) \in \delta \Rightarrow \xi' = \xi'', \omega' = \omega''$.

A non-deterministic chain $\delta$ over the subspace $\Omega$ can naturally be assigned the transition functions $\chi^c : \Omega \to 2^X$, $\rho^c : \Omega \to 2^X$, and $\hat{\rho}^c : \chi(\Omega) \to 2^X$ defined by

$$\chi^c := \chi|_\Omega, \quad \rho^c := \rho|_\Omega, \quad \hat{\rho}^c := \hat{\rho}_\Omega, \tag{16}$$

where $\chi$, $\rho$ and $\hat{\rho}_\Omega$ refer to the state machine $P$, as discussed in Section II-C. Then, *(i)* can be equivalently restated as $\chi^c(\omega') \cap \chi^c(\omega'') \neq \emptyset \Rightarrow \omega' = \omega''$, while *(ii)* is equivalent to $\hat{\rho}^c(\xi') \cap \hat{\rho}^c(\xi'') \neq \emptyset \Rightarrow \xi' = \xi''$. This leads us to the following statement.

*Proposition 1:* A transition relation $\delta$ over $\Omega$ is a non-deterministic chain if and only if $\chi^c : \Omega \to 2^X$ and $\hat{\rho}^c : \chi(\Omega) \to 2^X$ are absolutely injective set-valued maps.

Note that by definition (16) the functions $\chi^c$ and $\hat{\rho}^c$ take symbols as arguments. As a consequence of absolute injectivity of $\hat{\rho}^c$, for $w(t) \in \Omega$, from (4e) and (4f) it follows

$$\rho(w|_{[\tau,t]}) = \hat{\rho}^c \left[ \chi^c(w(t)) \cap \rho(w|_{[\tau,t-1]}) \right]. \tag{17}$$

This equation will shortly prove useful in designing the external space signal decomposition.

*Definition 3:* A state machine $P = (X, W, \Delta, X)$ is said to be *chain-decomposable* if $W$ and $\Delta$ can be partitioned as

$$W = \cup_{j=1}^r \Omega_j \quad \text{and} \quad \Delta = \cup_{j=1}^r \delta_j, \tag{18}$$

such that for each $j \in \{1, \ldots, r\}$, $\delta_j$ represents a non-deterministic chain over $\Omega_j$.

*Example 1:* Consider an I/S/- machine $P = (X, U \times Y, \Delta, X)$ with *singleton output maps*. Let $U = \{\mu_j; j = 1, \ldots, r\}$, and introduce the partitioning: $W = U \times Y = \cup_{j=1}^r \Omega_j$, where $\Omega_j := \mu_j \times Y$. This induces a partitioning of the transition relation: $\Delta = \cup_{j=1}^r \delta_j$. By definition, functions $f : X \times U \to 2^X$ and $h : X \times U \to Y$ exist, such that

$$(\xi, (\mu_j, \nu_j), \xi') \in \delta_j \Leftrightarrow \xi' \in f(\xi, \mu_j), \; \nu_j = h(\xi, \mu_j). \tag{19}$$

Then, each state $\xi \in X$ can be associated with a unique symbol pair $(\mu_j, \nu_j) \in \Omega_j$. Hence, *(i)* in Definition 2 is fulfilled. Define $\hat{\rho}_j^c : X \to 2^X$ as $\hat{\rho}_j^c(\xi) := f(\mu_j, \xi)$, and let it be *absolutely injective*. For instance, this is always true if the underyling vector field in case of a continuous time-driven system, refers to a Lipschitz continuous function on $X$. Hence, $\delta_j$ is a non-deterministic chain for all $j \in \{1, \ldots, r\}$. In particular, this conclusion holds if $P$ refers to an I/S/O machine and if $f(\mu_j, \xi)$ is injective for all $\mu_j \in U$; this case has been considered in [5]. ∎

Let $P$ be chain-decomposable and introduce *arbitrary* consistent signal space decompositions (8) for all subspaces $\Omega_j$, $j \in \{1, \ldots, r\}$ in (18):

$$\Omega_j \rightsquigarrow V_{j,1} \times \cdots \times V_{j,p}, \tag{20}$$

which leads to a decomposition of the external signal space $W \rightsquigarrow V_1 \times \cdots \times V_p$, as well, for instance by selecting

$$V_k = \cup_{j=1}^r V_{j,k}. \tag{21}$$

Next, we show that a decentralized scheme (see Fig. 1) built upon the distributed state machines $P_k$, $k \in \{1, \ldots, p\}$, which are induced by the underlying equivalence relations $\mathcal{A}_k$, will produce the exact outcomes of the monolithic state machine $P$. Referring to (16) and Proposition 1, it is important to keep in mind that the maps

$$\chi_j^c := \chi|_{\Omega_j} \quad \text{and} \quad \hat{\rho}_j^c := \hat{\rho}_{\Omega_j} \tag{22}$$

are, per construction, absolutely injective in their domains of definition. Now, fix an external signal $w|_{[\tau,t]}$, and consider any two different strings $w'|_{[\tau,t]}$ and $w''|_{[\tau,t]}$ from the subdomain $\mathcal{D}_\mathcal{A}(w|_{[\tau,t]})$ from the corresponding restriction domain $\mathcal{D}_\mathcal{A}(w|_{[\tau,t]})$ from (11). Note also that for any $\ell \in [\tau, t]$, all elements in $\mathcal{A}_k^{-1}(\mathcal{A}_k(w(\ell)))$ belong to the same subspace $\Omega_j$ for all $k \in \{1, \ldots, p\}$ and some $j \in \{1, \ldots, r\}$. Then, it follows from (17) and the injectivity of $\hat{\rho}_j^c$ that for any two different strings $w'|_{[\tau,t]}$ and $w''|_{[\tau,t]}$ from the subdomain $\mathcal{D}_\mathcal{A}(w|_{[\tau,t]})$:

$$\rho(w'|_{[\tau,t]}) \cap \rho(w''|_{[\tau,t]}) =$$
$$= \hat{\rho}_j^c [\rho(w'|_{[\tau,t-1]}) \cap \rho(w''|_{[\tau,t-1]}) \cap \chi_j^c(w'(t)) \cap \chi_j^c(w''(t))],$$

for some $j = j(t) \in \{1, \ldots, r\}$. Due to the injectivity of $\chi_j^c$, the argument on the right-hand side is an empty set, unless $w'(t) = w''(t) = w(t)$. In the latter case, the computation shrinks to the interval $[\tau, t-1]$:

$$\rho(w'|_{[\tau,t]}) \cap \rho(w''|_{[\tau,t]}) =$$
$$= \hat{\rho}_j^c(\chi_j^c(\omega(t)) \cap [\rho(w'|_{[\tau,t-1]}) \cap \rho(w''|_{[\tau,t-1]})]).$$

Yet a largest $\ell \in \{\tau, \ldots, t-1\}$ must exist, such that $w'(\ell) \neq w''(\ell)$. Then, after repeating this procedure $t - \ell$ times, the expression $\chi_j^c(w'(\ell)) \cap \chi_j^c(w''(\ell))$ appears in the argument of the function $\hat{\rho}_j^c$ for some $j = j(\ell) \in \{1, \ldots, r\}$, which is per construction empty, implying that the whole left hand side expression in the above equation must be empty, as well. In other words, $\rho : W^{[\tau,t]} \to 2^X$ results to be absolutely injective within the subdomain $\mathcal{D}_\mathcal{A}(w|_{[\tau,t]})$ inferred by the string $w|_{[\tau,t]}$. As a direct consequence:

$$\cap_{k=1}^p \rho_k(v_k|_{[\tau,t]}) = \cap_{k=1}^p \rho\left(\mathcal{A}_k^{-1}(\mathcal{A}_k(w|_{[\tau,t]}))\right)$$
$$= \rho\left(\cap_{k=1}^p \mathcal{A}_k^{-1}(\mathcal{A}_k(w|_{[\tau,t]}))\right)$$
$$= \rho\left(w|_{[\tau,t]}\right). \tag{23}$$

*Theorem 1:* Consider a chain-decomposable state machine $P = (X, W, \Delta, X)$ and a decentralized computation setting involving distributed state machines $P_k$ (see Fig. 1) induced by the equivalence relations $A_k$, $k \in \{1,\ldots,p\}$, resulting from the external signal space decomposition given by (20) and (21). Then, for any $w|_{[\tau,t]} \in W^{[\tau,t]}$:

$$\cap_{k=1}^{p} \chi_k(v_k|_{[\tau,t]}) = \chi(w|_{[\tau,t]}), \tag{24a}$$

$$\cap_{k=1}^{p} \rho_k(v_k|_{[\tau,t]}) = \rho(w|_{[\tau,t]}). \tag{24b}$$

*Proof:* The case $\tau = t$ in (24a) follows immediately from the absolute injectivity of a non-deterministic chain. In other cases, (24a) results after combining (24b), the recursive formula (4f) for $\chi$ and $\chi_k$, as well as the absolute injectivity of the maps $\chi_j^c$ for all $j \in \{1,\ldots,r\}$:

$$\begin{aligned}\cap_{k=1}^{p}\chi_k(v_k|_{[\tau,t]}) &= \cap_{k=1}^{p}\rho_k(v_k|_{[\tau,t-1]}) \cap \chi_k(v_k(t)) \\ &= \rho(w|_{[\tau,t-1]}) \cap \chi_j^c\left(\cap_{k=1}^{p}\mathcal{A}_k^{-1}(v_k(t))\right) \\ &= \chi(w|_{[\tau,t]}).\end{aligned} \tag{25}$$

Note that the derivations above put no further requirement but the consistency condition for the decomposition (20) of a signal subspace $\Omega_j$ corresponding to a nondeterministic chain $\delta_j$, $j \in \{1,\ldots,r\}$. This is a unique feature of nondeterministic chains. Yet, the opposite is not necessarily true: a transition relation $\delta$, which guarantees (25) under any decomposition (8), need not be a nondeterministic chain. ∎

*Example 2:* Consider the finite state machine in Fig. 3. Observe that it is chain decomposable. One way for partitioning its external signal space in accordance with (3) is $W = \Omega_1 \cup \Omega_2$ with $\Omega_1 = \{a_1, b_1, c_1, d_1\}$ and $\Omega_2 = \{a_2, b_2, c_2, d_2\}$.

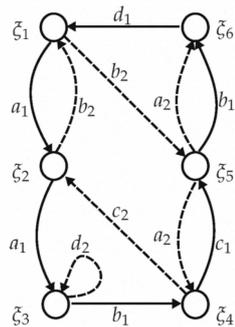

Fig. 3. Finite state machine.

As indicated by the solid and dashed lines, the corresponding transition relations $\delta_1$ and $\delta_2$ as defined by (15) are both non-deterministic chains. Note that $\delta_2$ is blocking and disconnected, which is a legal matter. According to the previous elaborations, any consistent decomposition (20) on subspaces $\Omega_1$ and $\Omega_2$ is legal. For instance, a particular is obtained from the selections

$V_{1,1} = \{\theta_1^1, \theta_1^2\}$ with $\theta_1^1 = L_1(\{a_1, b_1\}), \theta_1^2 = L_1(\{c_1, d_1\})$,
$V_{1,2} = \{\theta_2^1, \theta_2^2\}$ with $\theta_2^1 = L_2(\{a_1, c_1\}), \theta_2^2 = L_2(\{b_1, d_1\})$,
$V_{2,1} = \{\theta_1^3, \theta_1^4\}$ with $\theta_1^3 = L_1(\{a_2, b_2\}), \theta_1^4 = L_1(\{c_2, d_2\})$,
$V_{2,2} = \{\theta_2^3, \theta_2^4\}$ with $\theta_2^3 = L_2(\{a_2, c_2\}), \theta_2^4 = L_2(\{b_2, d_2\})$,

where $L_1$ and $L_2$ designate labeling functions. The inferred decomposition then reads $W \rightsquigarrow V_1 \times V_2$, where $V_1 = V_{1,1} \cup V_{2,1}$ and $V_2 = V_{1,2} \cup V_{2,2}$, leading to the distributed state machines $P_1 = (X, V_1, \Delta_1, X)$ and $P_2 = (X, V_2, \Delta_2, X)$. Now, consider a string, e.g. $w|_{[0,1]} = a_1 b_2$. The corresponding estimate of the monolithic machine is $\chi(a_1 b_2) = \xi_2$. The distributed machines measure accordingly the strings $v_1|_{[0,1]} = \theta_1^1 \theta_1^3$ and $v_2|_{[0,1]} = \theta_2^1 \theta_2^4$, providing the estimates $\chi_1(\theta_1^1 \theta_1^3) = \xi_2$ and $\chi_2(\theta_2^1 \theta_2^4) = \{\xi_2, \xi_3\}$, respectively. The decentralized estimate is thus given by $\chi(\theta_1^1 \theta_1^3) \cap \chi(\theta_2^1 \theta_2^4) = \xi_2$, which is exactly the same outcome obtained by the monolithic state machine $P = (X, W, \Delta, X)$. According to Theorem 1, this must hold for all strings accepted by the machine $P$. ∎

A state machine $P = (X, W, \Delta, X)$ involving a transition relation $\Delta$ such that $(\xi', \omega, \xi) \in \Delta$ and $(\xi'', \omega, \xi) \in \Delta$, with $\xi' \neq \xi''$ is not chain-decomposable due to the violation of the injectivity condition *(ii)* in Definition 2. On the other hand, condition *(i)* is always fulfilled by the trivial signal space partitioning (3) given by $\Omega_j = \omega_j$, $j \in \{1,\ldots,m\}$.

*Proposition 2:* A state machine $P = (X, W, \Delta, X)$ is chain decomposable if and only if $\hat{\rho}_W : W \to 2^X$ is absolutely injective.

### D. Generalized decomposition rules

Consider the state machine $P = (X, W, \Delta, X)$, and an equivalence relation $Q$ on the state space $X$, with $\pi_Q : X \to X/Q$ representing its canonical projection. Again, introduce a labeling for the equivalence classes $L_Z : X/Q \to Z$, and the map $\mathcal{Q} := L_Z \circ \pi_Q$:

$$\mathcal{Q} : X \to Z, \tag{26}$$

which represents the formal definition of an "aggregate" state space $Z$, denoted by $Z = L_Z(X/Q)$.

*Definition 4:* The quotient state machine $T = P/Q$ of $P = (X, W, \Delta, X)$ by the state space equivalence relation $Q$ is defined by $T = (Z, W, \Lambda, Z)$, with $Z = L_Z(X/Q)$ and $\Lambda \subseteq Z \times W \times Z$ given by

$$\Lambda = \{(\zeta, \omega, \zeta'); \exists \xi \in \mathcal{Q}^{-1}(\zeta), \xi' \in \mathcal{Q}^{-1}(\zeta'), (\xi, \omega, \xi') \in \Delta\} \tag{27}$$

Then, set-valued state functions $\chi' : W \to 2^Z$, $\rho' : W \to 2^Z$ and $\hat{\rho}'_{\Omega} : Z \to 2^Z$, $\Omega \subseteq W$, associated with $T$ exist such that

$$\chi' := \pi \circ \chi, \quad \rho' := \mathcal{Q} \circ \rho, \quad \hat{\rho}'_{\Omega} := \mathcal{Q} \circ \hat{\rho}_{\Omega}, \tag{28}$$

where $\chi$, $\rho$, $\hat{\rho}$ refer to the state machine $P$, as defined in Section II-C. The functions $\chi'$ and $\rho'$ are further naturally extended to mappings $W^{[\tau,t]} \to 2^Z$ in accordance with Section II-C. It is hereof important to state that by definition

$$\chi(\omega) \subseteq \mathcal{Q}^{-1} \circ \chi'(\omega) \tag{29}$$

and

$$\rho(w|_{[\tau,t]}) \subseteq \mathcal{Q}^{-1} \circ \rho'(w|_{[\tau,t]}) \tag{30}$$

hold for any $\omega \in W$ and any $w|_{[\tau,t]} \in W^{[\tau,t]}$, respectively.

Proposition 2 indicates that Theorem 1 does not apply for a state machine $P = (X, W, \Delta, X)$ if and only if the transition relation $\Delta$ implies non-injective state transition functions, i.e. if $\hat{\rho}_{\omega}(\xi') \cap \hat{\rho}_{\omega}(\xi'') \neq \emptyset$ for some $\omega \in W$ and $\xi' \neq \xi'' \in X$. Then, no non-deterministic chain $\delta$ in $P$ exists over some subspace $\Omega$ if $\omega \in \Omega$. To overcome this difficulty, in this section, we upgrade the algorithm for signal space decomposition by originating an equivalence relation $Q$ on the state space $X$ inducing a chain-decomposable quotient state machine $T = P/Q$. As we show shortly, this effectively leads to a further restriction of the region $\mathcal{D}_A(w|_{[\tau,t]})$ from (11), such that $\chi|_{\mathcal{D}_A(w|_{[\tau,t]})}$ and $\rho|_{\mathcal{D}_A(w|_{[\tau,t]})}$ are absolutely injective functions for any $w|_{[\tau,t]} \in W^{[\tau,t]}$.

*Lemma 1:* Consider a state machine $P = (X, W, \Delta, X)$ and let $Q$ be an equivalence relation defined by: $(\xi', \xi'') \in Q$ if $\xi \in X$ and $w|_{[\tau,t]} \in W^{[\tau,t]}$ exist, such that state space paths $(\xi', w|_{[\tau,t]}, \xi)$ and $(\xi'', w|_{[\tau,t]}, \xi)$ are present in $P$. Then, the quotient state machine $P/Q$ is chain-decomposable.

*Proof:* We provide the proof by contradiction. Let $T = P/Q = (Z, W, \Lambda, Z)$, and suppose that condition *(ii)* in Definition 2 is violated. Then, two different transitions $(\zeta', \omega, \zeta) \in \Lambda$ and $(\zeta'', \omega, \zeta) \in \Lambda$ must exist, with $\zeta' \neq \zeta''$. As a consequence, referring to Definition 4, $(\xi', \omega, \xi_1) \in \Delta$ and $(\xi'', \omega, \xi_2) \in \Delta$ exist, such that $\xi' \in \pi_Q^{-1}(\zeta')$, $\xi'' \in \pi_Q^{-1}(\zeta'')$ and $\xi_1, \xi_2 \in \pi_Q^{-1}(\zeta)$. Due to $\xi' \neq \xi''$, $\xi_1 \neq \xi_2$ ensues, otherwise $(\xi', \xi'') \in Q$ would entail $\zeta' = \zeta''$. As a consequence, with $(\xi_1, \xi_2) \in Q$, $w|_{[\tau,t]} \in W^{[\tau,t]}$ and $\xi \in X$ must exist, such that the paths $(\xi_1, w|_{[\tau,t]}, \xi)$ and $(\xi_2, w|_{[\tau,t]}, \xi)$ are present in $P$. But, $(\xi', \langle \omega, w|_{[\tau,t]} \rangle, \xi)$ and $(\xi'', \langle \omega, w|_{[\tau,t]} \rangle, \xi)$ are then present, as well, implying $(\xi', \xi'') \in Q$, which is in contradiction with the initial assumption of the proof. ∎

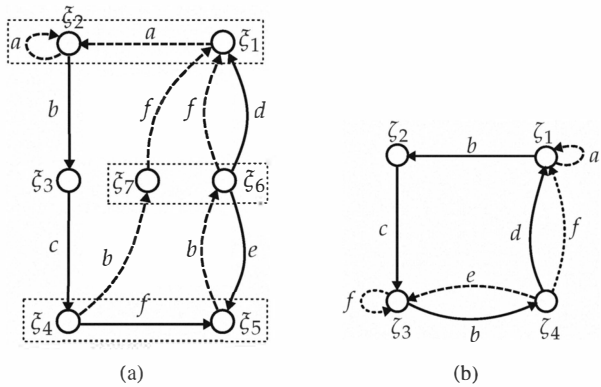

Fig. 4. Illustrating the generalized decomposition algorithm.

*Example 3:* For illustration purposes, consider $P = (X, W, \Delta, X)$ with $X = \{\xi_1, \xi_2, \xi_3, \xi_4, \xi_5, \xi_6, \xi_7\}$ and $W = \{a, b, c, d, e, f\}$, as depicted in Fig. 4(a). Fig. 4(b) shows a quotient state machine $T = (Z, W, \Lambda, Z)$ induced by the state space aggregation $Q$ leading to $Z = \{\zeta_1, \zeta_2, \zeta_3, \zeta_4\}$ in accordance with $\zeta_1 = L_Z(\{\xi_1, \xi_2\})$, $\zeta_2 = L_Z(\xi_3)$, $\zeta_3 = L_Z(\{\xi_4, \xi_5\})$ and $\zeta_4 = L_Z(\{\xi_6, \xi_7\})$. Note that $P$ is not chain decomposable due to the non-injectivity of functions $\hat{\rho}_a$ and $\hat{\rho}_f$, whereas $T$ is. As indicated in Fig. 4(b), the external signal space $W$ can be accordingly partitioned as $W = \Omega_1 \cup \Omega_2 \cup \Omega_3$, with $\Omega_1 = \{b, c, d\}$, $\Omega_2 = \{e, a\}$ and $\Omega_3 = \{f\}$. ∎

Now introduce a state space decomposition for the quotient state machine $T = (Z, W, \Lambda, Z)$ as suggested by the algorithm given by (20) and (21). Consider first any two symbols $\omega' \neq \omega'' \in W$. Then, using the fact that $\mathcal{Q}^{-1}$ and $\chi'$ are both absolutely injective functions, from (29) it follows

$$\chi(\omega') \cap \chi(\omega'') \subseteq \mathcal{Q}^{-1} \circ (\chi'(\omega') \cap \chi'(\omega''))$$

that is

$$\chi(\omega') \cap \chi(\omega'') = \emptyset. \tag{31}$$

Similarly, consider a $w|_{[\tau,t]} \in W^{[\tau,t]}$ and pick up any two different $w'|_{[\tau,t]} = w''|_{[\tau,t]}$ from the corresponding signal space restriction $\mathcal{D}_\mathcal{A}(w|_{[\tau,t]})$. Then, by using the fact that $\rho'|_{\mathcal{D}_\mathcal{A}(w|_{[\tau,t]})}$ is absolutely injective, from (30) it follows

$$\rho(w'|_{[\tau,t]}) \cap \rho(w''|_{[\tau,t]}) = \emptyset. \tag{32}$$

Hereof we conclude that as a consequence of our aggregation policy, absolutely injective mappings $\rho|_{\mathcal{D}_\mathcal{A}(w|_{[\tau,t]})}$ and $\chi|_{\mathcal{D}_\mathcal{A}(w|_{[\tau,t]})}$ are identified, which brings us to another main statement.

*Theorem 2:* Consider a state machine $P = (X, W, \Delta, X)$. A signal space decomposition (8) resulting from applying the algorithm (20) and (21) on the chain-decomposable quotient state machine $T = P/Q$, where the equivalence relation $Q$ is defined according to Lemma 1, leads invariably to exact computations (24a) and (24b) in a decentralized setup.

## IV. CONCLUSIONS

A general decentralized framework for set-valued state estimation and prediction for hybrid state machines has been discussed in this article. The outcome of the decentralized scheme is computed as the intersection of the outcomes provided by the individual distributed state machines. The latter are constructed by means of abstract "aggregation" maps that lead to a decomposition of the external signal space. In practice, such maps may refer to a set of sensors with different zooming and resolution capabilities. We show that, in the general case, decentralized schemes provide overapproximate estimate and prediction outcomes. However, using the concept of "non-deterministic chains", we are able to recover the exact computation outcomes with the help of constructive decomposition algorithms that take into account the structural properties of transition relation of the original state machine. The algorithms are systematically extended for transition relations that account for non-injectivity in the state transition function. In particular, we show that exact decentralized computation holds for all systems that assume an I/S/- realization with a singleton output map. Due to the "smaller" external signal spaces and transition relations associated with the individual distributed state machines, significant reduction in the overall space/time computational complexity may be expected. In this sense, a thorough complexity analysis on a case study can be found in [3]. Moreover, major advantages in terms of robustness and reliability are gained due to the redundancy of computation units. Our decentralized framework can be applied in different control contexts, including data fusion, failure detection & diagnosis, etc. Optimal construction of signal space decomposition leading to a minimization of the space/time complexity represents a natural extension to the work in this article.